\documentclass[twocolumn,superscriptaddress,aps,prl]{revtex4}
\usepackage{amsfonts}
\usepackage{amsmath}
\usepackage{amssymb}
\usepackage{graphicx}
\usepackage{color}
\usepackage{ulem}
\usepackage[colorlinks,urlcolor=cyan,citecolor=blue,linkcolor=magenta]{hyperref}

\begin{document}

\title{Floquet Engineering of a Dynamical $Z_{2}$ Lattice Gauge Field with Ultracold Atoms}
\author{Wei Zheng}
\email{zhengwei8796@gmail.com}
\affiliation{Hefei National Laboratory for Physical Sciences at the Microscale and
Department of Modern Physics, University of Science and Technology of China,
Hefei 230026, China }
\affiliation{CAS Center for Excellence in Quantum Information and Quantum Physics,
University of Science and Technology of China, Hefei 230026, China }
\author{Pengfei Zhang}
\email{pengfeizhang.physics@gmail.com}
\affiliation{Institute for Quantum Information and Matter,
California Institute of Technology, 
Pasadena, California 91125, USA}
\affiliation{Walter Burke Institute for Theoretical Physics,
California Institute of Technology, 
Pasadena, California 91125, USA}
\date{\today }

\begin{abstract}
In this paper, we propose that a simple model, in which fermions coupled to
a dynamical lattice gauge field, can be engineered via the Floquet approach.
The model possess both a independent Maxwell term and the local $Z_{2}$
gauge symmetry. Our proposal relies on a species dependent optical lattice,
and can be achieved in one, two or three dimension. By an unitary
transformation, this model can be mapped into a non-interacting composite
Fermion system with fluctuating back ground charge. With the help of this
composite Fermion picture, two characteristic observations are predicted.
One is the radio-frequency spectroscopy, which exhibits no dispersion in all
parameter regimes. Second is the the dynamical localization, which depends
on the structure of the initial states.
\end{abstract}

\maketitle

The Gauge field theory is one of the most important footing stone of the
modern physics. On one hand, the gauge theory successfully described the
fundamental interactions within the standard model of high energy physics.
On another hand, lattice gauge theory can also emerges from the low-energy
physics of the strongly correlated systems~\cite{Rev@Kogut.1979}. For example, the $U(1)$ lattice
gauge theory can effectively describe the collective fluctuations of the
frustrated magnetism~\cite%
{fru_mag@Balents.2010,fru_mag@Read.1991,fru_mag@Sachdev.1991}, quantum spin
liquid~\cite{SL@Anderson.1988,SL@Read.1988,SL@Affleck.1988,SL@Fisher.2000},
quantum dimer model~\cite{QDM@Kivelson.1987,QDM@Fradkin.2001}. In fact,
dynamical gauge fields entre the theory naturally in the so-called
slave-particle decomposition, whereby the original particles are
fractionalized into slave degrees of freedom. The relative phase between the
slave particles introduces the local gauge invariance. Understanding the
real time dynamics of lattice gauge fields is a notable challenge due to the
limit of the classical computational methods. This stimulates the efforts on
the quantum simulation of the gauge fields in various engineered systems~
\cite{U1EXP@Zoller.2016,U1EXP@Yuan.2020,U1EXP@Jendrzejewski.2020}.
\cite{LGT_REV@Wiese.2013,LGT_CAP@Demler.2018,LGT@Grover.2016,Z2@Cirac.2017}

In cold atom system, static artificial gauge fields has been simulated in
recent years, including electrical field~\cite{Electric@Spielman.2011}, magnetic field~\cite{Magnetic@Spielman.2009}, and even spin-orbit
couplings~\cite{SOC@Spielman.2011,2DSOC@Chen.Shuai.2016}. However the configuration of those gauge fields is fixed by the
external laser or the system geometry. Therefore these fields have no
dynamics, and can not mediate the interaction between atoms. Recently,
Chicago and ETH group have realized the density-dependent gauge field by
periodic modulation of the optical lattice or the interaction~\cite%
{DDGF@Chin.2018,DDGF@Esslinger.2019}; while Stanford group has realized a
cavity-induced one dimensional spin-orbit-coupled Bose-Einstein condensate~%
\cite{Cavity_SOC@Lev.2019}. In these experiments, the artificial gauge
fields are no longer static, but can fluctuate with time. However the local
gauge invariance, the key ingredient of the gauge theory, is still missing.
In 2019, Bloch's group successfully simulated the dynamical $Z_{2}$ lattice
gauge field in a double well potential~\cite{Z2EXP@Bloch.2019}, where the coupled dynamics of a single boson and the gauge field are observed.

In this paper, inspired by the Bloch's experiment, we propose that a
simplest model, in which fermions interacting with a dynamical lattice gauge
field with local $Z_{2}$ invariant, can be engineered via the Floquet
approach. Our setup can be applied in one, two or three dimension. By an
unitary transformation, this model can be mapped into a non-interacting
composite Fermion system with fluctuating back ground charge. With the help
of this composite Fermion picture, two characteristic observations are
predicted. One is that the radio-frequency spectrum of the Fermions exhibits no dispersion in all
parameter regimes, which reflects the local symmetry of the model. Second is
the dynamical localization of the Fermions, which depends on the structure of the initial
states.

\begin{figure}[tbp]
\includegraphics[width=3.2in]{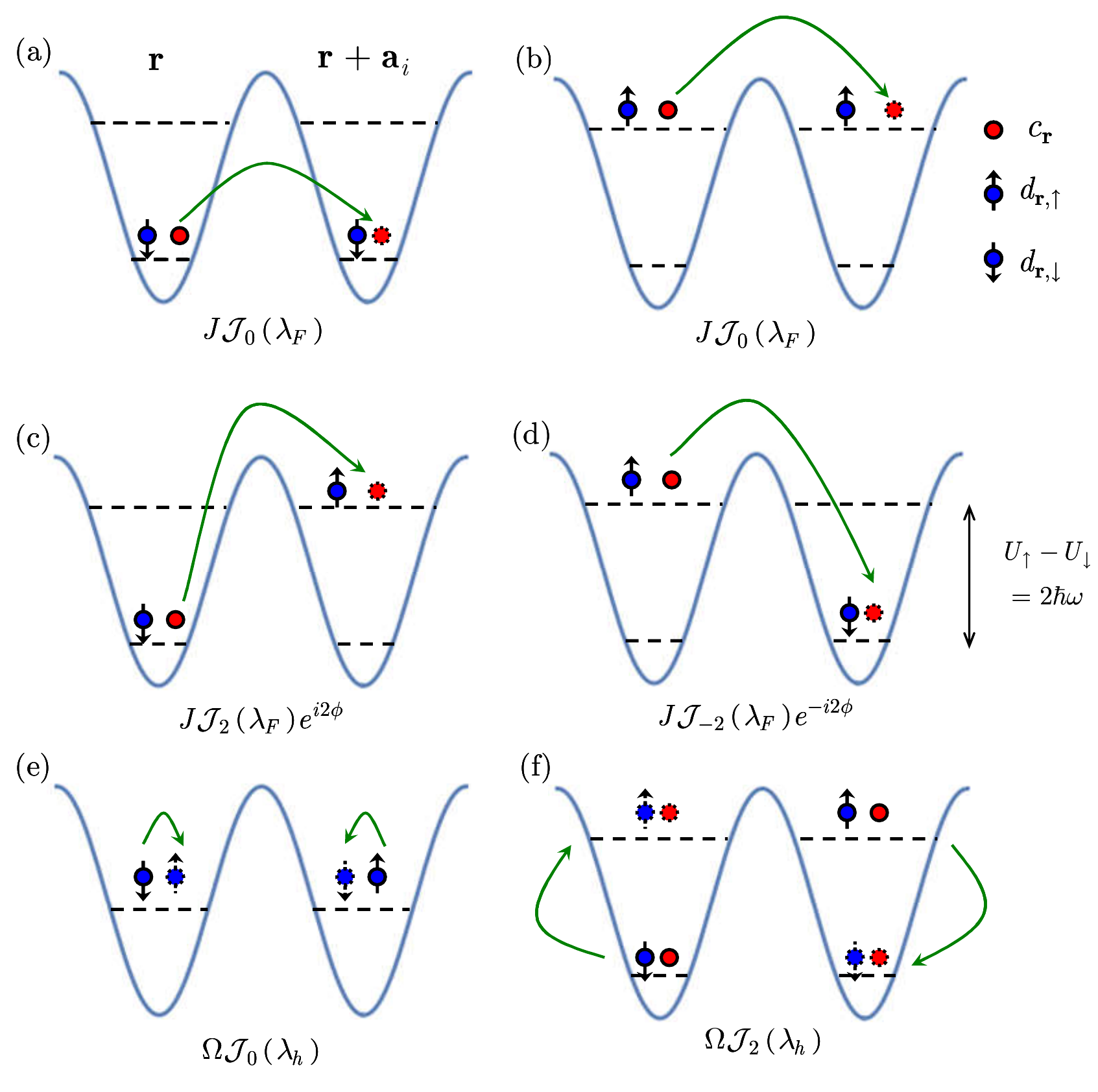}
\caption{The physical processes keeped in effective Hamiltonian. Red circles
represent the spinless $\hat{c}$-fermions hopping in the lattice; while blue
circles with arrow represent the spinful localized $\hat{d}$-fermions.
(a)(b) The renormalized hopping of $\hat{c}$-fermions in the case of same
spins polarize of $\hat{d}$-fermions between the neighbor sites. (c)(d) The
shaking assistant hopping of $\hat{c}$-fermions in the case of opposite spin
polarizations of $\hat{d}$-fermions. (e) The renormalized spin flip of $\hat{%
d}$-fermions in the absence of $\hat{c}$-fermions. (f) The shaking assistant
spin flip of $\hat{d}$-fermions in the presence of $\hat{c}$-fermions.}
\label{fig1}
\end{figure}

\textit{Experiment setup}. We consider two kinds of atoms, one is mobile
spinless fermion noted by $\hat{c}_{\mathbf{r}}$, which is used to simulate
the matter field; while another is localized spin-$1/2$ fermions noted by $%
\hat{d}_{\mathbf{r,\uparrow (\downarrow )}}$ to simulate the gauge field.
The fermions are trapped in a species dependent cubic optical lattice. By
carefully tuning the optical lattice, one can highly suppress the hopping of
the $\hat{d}$-fermions; while keeping the hopping of the $\hat{c}$-fermions.
An radio-frequency field is applied to coupled spin up and spin down states
of the $\hat{d}$-fermions. The tight-binding Hamiltonian reads $\hat{H}(t)=%
\hat{H}_{0}+\hat{H}_{\mathrm{I}}+\hat{V}(t)$, where
\begin{eqnarray}
H_{0} &=&J\sum\limits_{\mathbf{r,}i}\hat{c}_{\mathbf{r}+\mathbf{a}%
_{i}}^{\dag }\hat{c}_{\mathbf{r}}+\Omega \sum\limits_{\mathbf{r}}\hat{d}_{%
\mathbf{r,\uparrow }}^{\dag }\hat{d}_{\mathbf{r,\downarrow }}+h.c., \\
\hat{H}_{\mathrm{I}} &=&\sum\limits_{\mathbf{r}}(U_{\uparrow }\hat{n}_{%
\mathbf{r,\uparrow }}^{d}\hat{n}_{\mathbf{r}}^{c}+U_{\downarrow }\hat{n}_{%
\mathbf{r,\downarrow }}^{d}\hat{n}_{\mathbf{r}}^{c}), \\
\hat{V}(t) &=&\sum\limits_{\mathbf{r}}\left[ \mathbf{F}(t)\mathbf{\cdot r}%
\hat{n}_{\mathbf{r}}^{c}+h(t)(\hat{n}_{\mathbf{r,\uparrow }}^{d}-\hat{n}_{%
\mathbf{r,\downarrow }}^{d})\right] .
\end{eqnarray}%
Here $\mathbf{a}_{i}=a\mathbf{e}_{i}$, is the basic vector of the cubic
lattice, $U_{\uparrow (\downarrow )}$ is the on-site interaction between the
$\hat{c}$-fermions and spin up (down) $\hat{d}$-fermions. The interaction
between the spin-up and spin-down $\hat{d}$-fermions is irrelevant in this
problem, since we constrain the particle number of $\hat{d}$-fermion at each
site to be unit, $\hat{n}_{\mathbf{r,\uparrow }}^{d}+\hat{n}_{\mathbf{%
r,\downarrow }}^{d}=1$. There are two periodic driving term. One is an
oscillating Zeeman field, $h(t)=h\cos (\omega t)$, which is applied on the $%
\hat{d}$-fermions. Another is an periodic driving force, $\mathbf{F}%
(t)=F\cos (\omega t+\phi )\sum\limits_{i=1}^{3}\mathbf{e}_{i}$, which can be
realized by linear polarized shaking of the optical lattice. The two driving
terms have the same frequency with a phase shift $\phi $. We choose the
driving frequency to be two-photon resonance with the interaction difference
$2\hbar \omega =U_{\uparrow }-U_{\downarrow }$, and the phase shift to be $%
\phi =\pi /2$.

To deal with this time periodic driven quantum system, we first make a
rotating transformation, $\hat{R}(t)=\exp \left\{ -\frac{i}{\hbar }%
\int\nolimits_{0}^{t}d\tau \lbrack \hat{H}_{\mathrm{I}}+\hat{V}(\tau
)]\right\} ,$ to eliminate the interaction and the driving terms. One
obtains the Hamiltonian in the rotating frame as
\begin{eqnarray}
\hat{H}_{\mathrm{rw}}(t) &=&J\sum\limits_{\mathbf{r,}i}\hat{c}_{\mathbf{r}+%
\mathbf{a}_{i}}^{\dag }\hat{P}_{\mathbf{r}+\mathbf{a}_{i}}^{d}\hat{P}_{%
\mathbf{r}}^{d}\hat{c}_{\mathbf{r}}e^{i\frac{Fa}{\hbar \omega }\sin (\omega
t)}+  \notag \\
&&\Omega \sum\limits_{\mathbf{r}}\hat{d}_{\mathbf{r,\uparrow }}^{\dag }\hat{P%
}_{\mathbf{r,\uparrow }}^{c}\hat{P}_{\mathbf{r,\uparrow }}^{c}\hat{d}_{%
\mathbf{r,\downarrow }}e^{i\frac{2h}{\hbar \omega }\cos (\omega t)}+h.c.,
\end{eqnarray}%
where $\hat{P}_{\mathbf{r}}^{d}=\hat{n}_{\mathbf{r,\uparrow }%
}^{d}e^{-iU_{\uparrow }t/\hbar }+\hat{n}_{\mathbf{r,\downarrow }%
}^{d}e^{-iU_{\downarrow }t/\hbar }$, and $\hat{P}_{\mathbf{r,\uparrow }%
\left( \mathbf{\downarrow }\right) }^{c}=(1-\hat{n}_{\mathbf{r}}^{c})+\hat{n}%
_{\mathbf{r}}^{c}e^{-iU_{\mathbf{\uparrow }\left( \downarrow \right)
}t/\hbar }$ are oscillating phase factors, depending on the occupation of $%
\hat{d}$-fermion and $\hat{c}$-fermions respectively.

Then we apply the high frequency expansion to seek the time independent
effective Hamiltonian as $\hat{H}_{\mathrm{eff}}=\hat{H}_{\mathrm{rw}%
}^{(0)}+\sum_{m=1}^{\infty }\frac{\left[ \hat{H}_{\mathrm{rw}}^{(m)},\hat{H}%
_{\mathrm{rw}}^{(-m)}\right] }{m\hbar \omega }+\cdots $, where $\hat{H}_{%
\mathrm{rw}}^{(m)}=\frac{1}{T}\int\nolimits_{0}^{T}dt\hat{H}_{\mathrm{rw}%
}(t)e^{-im\omega t}$ is the Fourier component of the Hamiltonian. Here we
consider the large frequency limit, such that one can only keep to zeroth
order, obtaining%
\begin{eqnarray}
\hat{H}_{\mathrm{eff}} &\approx &J\sum\limits_{\mathbf{r,}i}\left[ \mathcal{J%
}_{0}(\lambda _{F})(\hat{n}_{\mathbf{r}+\mathbf{a}_{i}\mathbf{,\uparrow }%
}^{d}\hat{n}_{\mathbf{r,\uparrow }}^{d}+\hat{n}_{\mathbf{r}+\mathbf{a}_{i}%
\mathbf{,\downarrow }}^{d}\hat{n}_{\mathbf{r,\downarrow }}^{d})\right.
\notag \\
&&\left. -\mathcal{J}_{2}(\lambda _{F})(\hat{n}_{\mathbf{r}+\mathbf{a}_{i}%
\mathbf{,\downarrow }}^{d}\hat{n}_{\mathbf{r,\uparrow }}^{d}+\hat{n}_{%
\mathbf{r}+\mathbf{a}_{i}\mathbf{,\uparrow }}^{d}\hat{n}_{\mathbf{%
r,\downarrow }}^{d})\right] \hat{c}_{\mathbf{r}+\mathbf{a}_{i}}^{\dag }\hat{c%
}_{\mathbf{r}}  \notag \\
&&+\Omega \sum\limits_{\mathbf{r}}\left[ \mathcal{J}_{0}(\lambda _{h})(1-%
\hat{n}_{\mathbf{r}}^{c})\right.  \notag \\
&&\left. +\mathcal{J}_{2}(\lambda _{h})\hat{n}_{\mathbf{r}}^{c}\right] \hat{d%
}_{\mathbf{r,\uparrow }}^{\dag }\hat{d}_{\mathbf{r,\downarrow }}+h.c.,
\end{eqnarray}%
where $\lambda _{F}=\frac{Fa}{\hbar \omega }$, $\lambda _{h}=\frac{2h}{\hbar
\omega }$, and $\mathcal{J}_{m}\left( \lambda \right) $ is the Bessel
function. The corresponding physical processes are illustrated in Fig.\ref%
{fig1}. The lattice shaking largely modifies the hopping of the $\hat{c}$%
-fermions so that it is depended on the spin configuration of the localized $%
\hat{d}$-fermions. If the spins polarize to the same direction in the
nearest neighbor sites, there is no energy offset for the $\hat{c}$%
-fermions. So the lattice shaking will only renormalized the bare hopping to
$J\mathcal{J}_{0}(\lambda _{F})$, see Fig.\ref{fig1}(a)(b). If the spin
polarizations are opposite to each other, the energy imbalance is $\pm
\left( U_{\uparrow }-U_{\downarrow }\right) $. Then the $\hat{c}$-fermions
will absorb(emit) two energy quanta from(to) the shaking to hope. So the
amplitude is $J\mathcal{J}_{2}(\lambda _{F})$. Moreover, the phase of these
two energy quanta is imprinted to the hopping, see Fig.\ref{fig1}(c)(d).
Similar to the hopping process, the spin flip of the $\hat{d}$-fermions is
also modified by oscillating Zeeman field. It is dependent on the charge
density of the $\hat{c}$-fermions. If $n_{\mathbf{r}}^{c}=0$, the Rabi
frequency is merely renormalized by the oscillating Zeeman field, see Fig.%
\ref{fig1}(e). If $n_{\mathbf{r}}^{c}=1$, two energy quanta is
absorbed(emitted) to assistant the local spin flip, see Fig.\ref{fig1}(f).

\begin{figure}[tbp]
\includegraphics[width=3.3in]{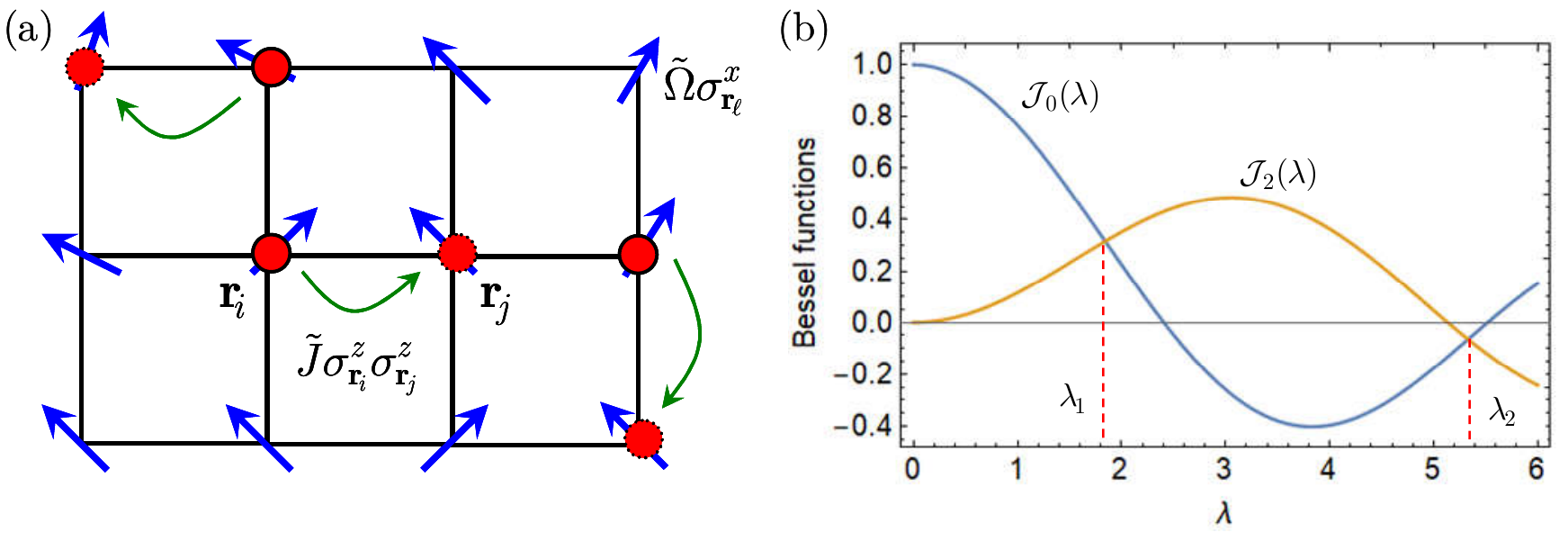}
\caption{(a) Illustration of the $Z_{2}$ lattice gauge theory given by
Hamiltonian (\protect\ref{Ham_LGT}). Local spins live on each lattice site.
Spinless $\hat{c}$-fermions can hope between the lattice sites. The hopping
amplitude depends on the configuration of the local spins. (b) Fine tuning
of modulations. The solutions of $\mathcal{J}_{0}\left( \protect\lambda %
\right) =\mathcal{J}_{2}\left( \protect\lambda \right) $ are $\protect%
\lambda _{1}=1.84118$ and $\protect\lambda _{2}=5.33144$. }
\label{fig2}
\end{figure}

We fine tune the parameter as $\mathcal{J}_{0}\left( \lambda _{F}\right) =%
\mathcal{J}_{2}\left( \lambda _{F}\right) $ and $\mathcal{J}_{0}\left(
\lambda _{h}\right) =\mathcal{J}_{2}\left( \lambda _{h}\right) $. This can
be achieved by choosing $\lambda _{F},\lambda _{h}=1.84118$ or $5.33144$,
see Fig.\ref{fig2}. Then we use the Pauli matrix to represent the spin
degree of freedom of the local $\hat{d}$-fermions as $\sigma _{\mathbf{r}%
}^{z}=\hat{n}_{\mathbf{r,\uparrow }}^{d}-\hat{n}_{\mathbf{r,\downarrow }%
}^{d} $, $\sigma _{\mathbf{r}}^{x}=\hat{d}_{\mathbf{r,\uparrow }}^{\dag }%
\hat{d}_{\mathbf{r,\downarrow }}+\hat{d}_{\mathbf{r,\downarrow }}^{\dag }%
\hat{d}_{\mathbf{r,\uparrow }}$, and $\sigma _{\mathbf{r}}^{y}=-i(\hat{d}_{%
\mathbf{r,\uparrow }}^{\dag }\hat{d}_{\mathbf{r,\downarrow }}-\hat{d}_{%
\mathbf{r,\downarrow }}^{\dag }\hat{d}_{\mathbf{r,\uparrow }})$. So the
Hamiltonian can be rewritten into%
\begin{equation}
\hat{H}_{\mathrm{eff}}=\tilde{J}\sum\limits_{\mathbf{r,}i}\sigma _{\mathbf{r}%
+\mathbf{a}_{i}}^{z}\sigma _{\mathbf{r}}^{z}(\hat{c}_{\mathbf{r}+\mathbf{a}%
_{i}}^{\dag }\hat{c}_{\mathbf{r}}+h.c.)+\sum\limits_{\mathbf{r}}\tilde{\Omega%
}\sigma _{\mathbf{r}}^{x},  \label{Ham_LGT}
\end{equation}%
where $\tilde{J}=J\mathcal{J}_{0}\left( \lambda _{F}\right) $ and $\tilde{%
\Omega}=\Omega \mathcal{J}_{0}\left( \lambda _{h}\right) $. This Hamiltonian
describes a dynamic $Z_{2}$ lattice gauge field coupled to spinless
fermions. We note that the fermion hopping phase factor $e^{i\theta }$,
which is $\sigma _{\mathbf{r}+\mathbf{a}_{i}}^{z}\sigma _{\mathbf{r}}^{z}$,
could be $\pm 1$. That means the Peierls phase $\theta $ is either $0$ or $%
\pi $, rather than a continuous value in the $U(1)$ gauge theory. Moreover, $%
\sigma _{\mathbf{r}+\mathbf{a}_{i}}^{z}\sigma _{\mathbf{r}}^{z}$ does not
commute with the Maxwell term, $\tilde{\Omega}\sum\nolimits_{\mathbf{r}%
}\sigma _{\mathbf{r}}^{x}$, in the Hamiltonian. As a result, unlike the
static gauge field, Peierls phase in this lattice will fluctuate in the real
time dynamics.

The Hamiltonian (\ref{Ham_LGT}) can also be dualed to the standard $Z_{2}$
lattice gauge theory, where gauge fields $\tau $ are defined on the link,
with coupling terms $\hat{c}_{\mathbf{r}+\mathbf{a}_{i}}^{\dag }\hat{c}_{%
\mathbf{r}}\tau _{\mathbf{r}}^{z}$, star terms $\prod_{+}\tau _{\mathbf{%
r+a_{i}}/2}^{x}$ and flux terms $\prod_{\square }\tau _{\mathbf{r}}^{z}$ in
the Hamiltonian. We could define the duality $\tau _{\mathbf{r+a_{i}}%
/2}^{z}=\sigma _{\mathbf{r}+\mathbf{a}_{i}}^{z}\sigma _{\mathbf{r}}^{z}$.
The the first term in (\ref{Ham_LGT}) becomes the standard coupling term and
the second term becomes the star term. Moreover, the definition imposes that
$\prod_{\square }\tau _{\mathbf{r}}^{z}=1$, which is equivalent to having a
large flux term.

\textit{Symmetry. }Unlike the density-dependent gauge fields without local
gauge symmetry \cite{DDGF@Chin.2018,DDGF@Esslinger.2019,Cavity_SOC@Lev.2019}%
, our model possesses a local $Z_{2}$ gauge symmetry at each site. The gauge
transformation operator is given by
\begin{equation}
\hat{G}_{\mathbf{r}}=e^{i\pi \hat{n}_{\mathbf{r}}^{c}}\sigma _{\mathbf{r}%
}^{x}.
\end{equation}%
It is easy to check $[\hat{G}_{\mathbf{r}},\hat{H}_{\mathrm{eff}}]=0$, and $[%
\hat{G}_{\mathbf{r}_{2}},\hat{G}_{\mathbf{r}_{1}}]=0$ for any given site.
That means the eigenstates of the Hamiltonian is also the eigenstates of $%
\hat{G}_{\mathbf{r}}$, $\hat{G}_{\mathbf{r}}\left\vert \psi \left( \mathbf{g}%
\right) \right\rangle =g_{\mathbf{r}}\left\vert \psi \left( \mathbf{g}%
\right) \right\rangle $, where vector $\mathbf{g}=\left( g_{\mathbf{r}%
_{1}},g_{\mathbf{r}_{2}},\cdots \right) $. Here the eigen value $g_{\mathbf{r%
}}=\pm 1$. This separates the Hilbert space into different subsectors. In
each subsector, $\hat{G}_{\mathbf{r}}$ is a good quantum number, giving $%
e^{i\pi \hat{n}_{\mathbf{r}}^{c}}g_{\mathbf{r}}=\sigma _{\mathbf{r}}^{x}$,
which is nothing but the $Z_{2}$ Gauss's law similar to the electrodynamics.
The $e^{i\pi \hat{n}_{\mathbf{r}}^{c}}$ is the $Z_{2}$ charge carried by
matter field, the $g_{\mathbf{r}}$ is the local black ground charge, and $%
\sigma _{\mathbf{r}}^{x}$ is the electrical field. In real life, we are
constrained in one subspace of the $U(1)$ gauge theory, where the background
charge is zero. However in our synthetic world, we can prepare the initial
states at different subspaces, or even the superposition of several
subsectors. That leads richer dynamics of the gauge field.

Making a unitary transformation\cite{Z2LGT@Maciejko.2017}, $U=e^{i\frac{\pi
}{2}\sum\nolimits_{\mathbf{r}}\sigma _{\mathbf{r}}^{z}\hat{n}_{\mathbf{r}%
}^{c}}$, we find that the matter field and the gauge field are bound,
\begin{eqnarray}
\hat{c}_{\mathbf{r}} &\longrightarrow &-i\sigma _{\mathbf{r}}^{z}\hat{c}_{%
\mathbf{r}},  \label{bound} \\
\sigma _{\mathbf{r}}^{z} &\longrightarrow &\sigma _{\mathbf{r}}^{z}, \\
\sigma _{\mathbf{r}}^{x} &\longrightarrow &\sigma _{\mathbf{r}}^{x}-2\sigma
_{\mathbf{r}}^{x}\hat{n}_{\mathbf{r}}.
\end{eqnarray}%
As a results, the Hamiltonian and $\hat{G}_{\mathbf{r}}$ transforms into
\begin{eqnarray}
\hat{H}^{\prime } &=&\tilde{J}\sum\limits_{\mathbf{r},i}(\hat{c}_{\mathbf{r}+%
\mathbf{a}_{i}}^{\dag }\hat{c}_{\mathbf{r}}+h.c.)+\tilde{\Omega}\sum\limits_{%
\mathbf{r}}\sigma _{\mathbf{r}}^{x}\left( 1-2\hat{n}_{\mathbf{r}}\right) ,
\notag \\
\hat{G}_{\mathbf{r}}^{\prime } &=&\sigma _{\mathbf{r}}^{x},
\end{eqnarray}%
Note that after transformation, Hamiltonian explicitly includes local gauge
transformation operator. In one particular subsector, $\hat{G}_{\mathbf{r}%
}^{\prime }$ is a good quantum number, and can be replaced by its eigen
values. So Hamiltonian in this subsector reads
\begin{eqnarray}
\hat{H}_{\mathrm{sub}}\left( \mathbf{g}\right)  &=&\tilde{J}\sum\limits_{%
\mathbf{r,}i}(\hat{c}_{\mathbf{r}+\mathbf{a}_{i}}^{\dag }\hat{c}_{\mathbf{r}%
}+h.c.)  \notag \\
&&-2\tilde{\Omega}\sum\limits_{\mathbf{r}}g_{\mathbf{r}}\hat{n}_{\mathbf{r}%
}^{c}+\tilde{\Omega}\sum\limits_{\mathbf{r}}g_{\mathbf{r}}
\end{eqnarray}%
By constrained in one subsector, this lattice gauge model is mapped into
non-interacting fermions moving in the potential of the background charges.
These fermions are so-called "Orthogonal" fermions~\cite%
{OF@Sigrist.2010,OF@Senthil.2012,OF@Sachdev.2019,OF@Meng.2020}, which are
composite of local spins and $\hat{c}$-fermions. This mapping is similar to
the integral out of the longitudinal electrical field to obtain the Coulomb
interaction in the Maxwell theory. In our model, there is no Coulomb
interaction between the particles, but only the coupling between matter
field and the background charges.

\begin{figure}[tbp]
\includegraphics[width=3.3in]{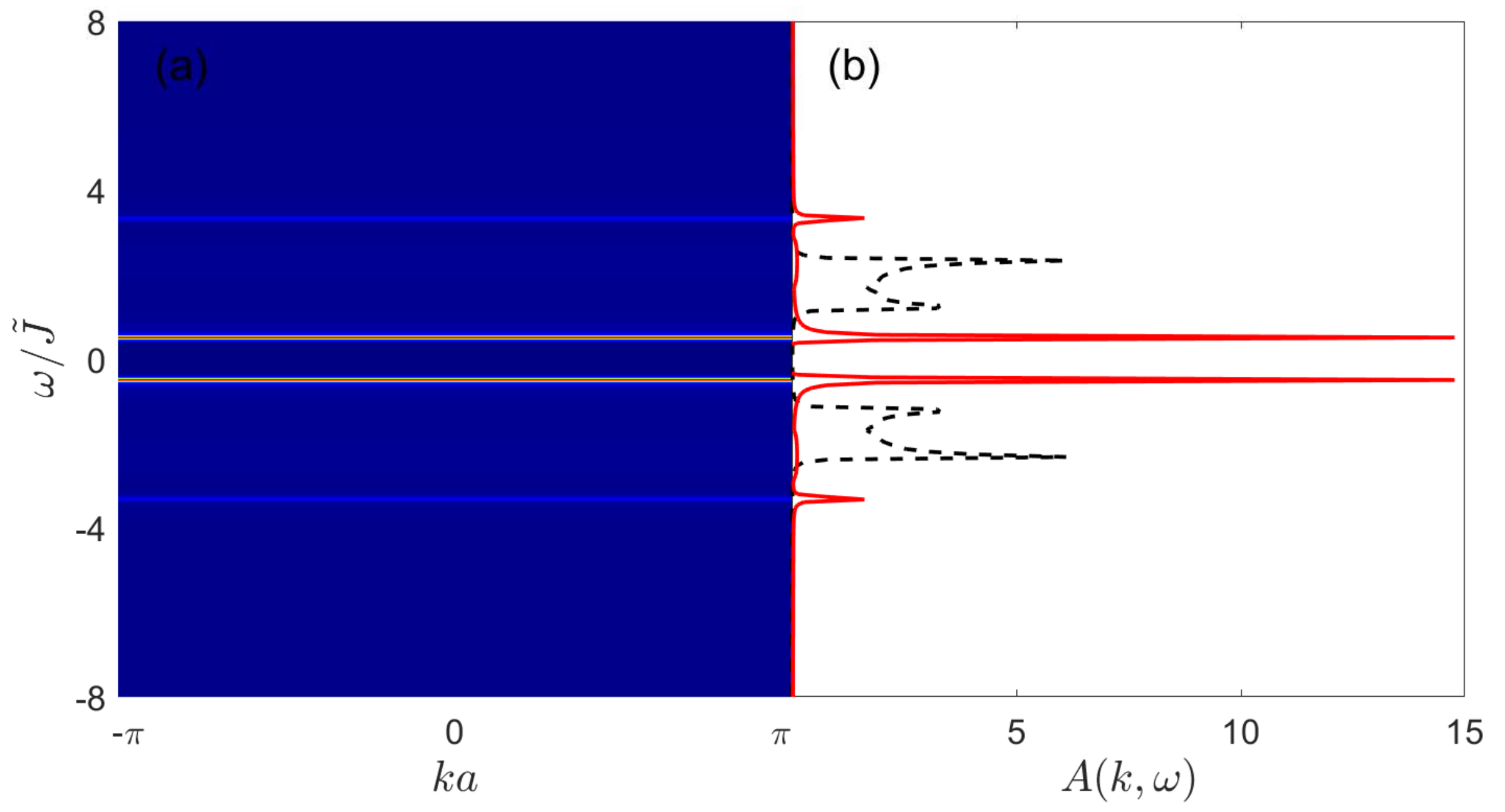}
\caption{(a) Momentum resolved radio-frequency spectrum function for a
half-filled one dimensional lattice at zero-temperature. (b) Radio-frequency
spectrum function at a given momentum. The dashed line is the
single-particle density-of-state in a spin density wave back ground charge
distribution. The calculation is done in a lattice with length $100$, and $%
\tilde{\Omega}/\tilde{J}=0.6$. }
\label{fig3}
\end{figure}

At the half-filling, we numerically found the ground state as $\left\vert
\psi _{\mathrm{gs}}\right\rangle =\hat{U}^{\dag }\left\vert \varphi _{%
\mathrm{CDW}}\right\rangle \otimes \left\vert \mathbf{g}^{\mathrm{SDW}%
}\right\rangle $ in one dimension. The local spins form a spin density wave
(SDW), $g_{\mathbf{r}}^{\mathrm{SDW}}=\left( -1\right)
^{\sum\nolimits_{i=1}^{d}r_{i}}$. The "Orthogonal" fermions form a charge
density wave (CDW), $\left\vert \varphi _{\mathrm{CDW}}\right\rangle $. This
ground state can be generalized to two and three dimension. It can be
understood in two limits. First if $\tilde{J}\rightarrow 0$, the ground
states have $\left\langle \hat{n}_{\mathbf{r}}\right\rangle =1(0)$, as $g_{%
\mathbf{r}}=+1(-1)$ to lower the on-site energy. There are innumerable
states possessing the same energy, but for a finite $\tilde{J}$, the SDW
configuration will minimize the kinetic energy at the half-filling. Second,
at $\tilde{\Omega}\rightarrow 0$, the ground state is a Fermi sea of the
"Orthogonal" fermions. When openning $\tilde{\Omega}$, the SDW will
scattering the "Orthogonal" fermions by a momentum $\mathbf{Q}%
=\sum\nolimits_{i=1}^{d}\frac{\pi }{a}\mathbf{e}_{i}$. At the half-filling,
it will mach the Fermi surface nesting to open a gap, leading a CDW.

\textit{Observations}. The first observation is the radio-frequency
spectroscopy. Considering the retarder Green's function of the $\hat{c}$%
-fermions, $iG_{\mathbf{r}_{2},\mathbf{r}_{1}}^{\mathrm{R}}\left( t\right)
=\Theta \left( t\right) \left\langle \{\hat{c}_{\mathbf{r}_{2}}(t),\hat{c}_{%
\mathbf{r}_{1}}^{\dag }(0)\}\right\rangle $. Transforming into the "Orthogonal"
fermion picture, we have
\begin{equation}
G_{\mathbf{r}_{2},\mathbf{r}_{1}}^{\mathrm{R}}\left( t\right) =\delta _{%
\mathbf{r}_{2}\mathbf{,r}_{1}}\sum\limits_{\mathbf{g}}S_{\mathbf{r}%
_{1}}\left( t,\mathbf{g}\right)
\end{equation}%
Here $iS_{\mathbf{r}}\left( t,\mathbf{g}\right) =\Theta \left( t\right)
\sum\limits_{n}\frac{e^{-\beta E_{n}\left( \mathbf{g}\right) }}{Z}%
\left\langle \varphi _{n}\left( \mathbf{g}\right) \right\vert \hat{c}_{%
\mathbf{r}}\times e^{-\frac{i\left[ H_{\mathrm{sub}}\left( \mathbf{g}%
^{\prime }\right) -E_{n}\left( \mathbf{g}\right) \right] t}{\hbar }}\hat{c}_{%
\mathbf{r}}^{\dag }\left\vert \varphi _{n}\left( \mathbf{g}\right)
\right\rangle +\left( \hat{c}_{\mathbf{r}}\leftrightarrow \hat{c}_{\mathbf{r}%
}^{\dag },t\leftrightarrow -t\right) $ where $E_{n}\left( \mathbf{g}\right) $
is the eigen energy, $\left\vert \varphi _{n}\left( \mathbf{g}\right)
\right\rangle $ the eigen state of "Orthogonal" fermion at the given background
charge $\mathbf{g}$. The background charge distribution $\mathbf{g}^{\prime
} $ is opposite from $\mathbf{g}$ at site $\mathbf{r}$, $\left\vert \mathbf{g%
}^{\prime }\right\rangle =\sigma _{\mathbf{r}}^{z}\left\vert \mathbf{g}%
\right\rangle $. The prefactor $\delta _{\mathbf{r}_{2}\mathbf{,r}_{1}}$
indicates the single particle(hole) excitations can not propagate, since
they are bound with the local spins (\ref{bound}). As a results the momentum
resolved radio-frequency spectroscopy, which measured the imaginary part of
the retarded Green's function, shows no dispersion at any given temperature,
$A\left( \mathbf{k},\omega \right) =-2\mathrm{Im}G_{\mathbf{0}}^{\mathrm{R}%
}\left( \omega \right) $. In addition, adding(removing) the $\hat{c}$%
-fermions to(from) the system, will flip the local spins, which will create
an local impurity potential for the "Orthogonal" fermions, this process is
similar to the X-Ray absorption or emission in the solid materials~\cite%
{OC@Anderson.1967,OC@Mahan.1967,OC@Nozieres1,OC@Nozieres2,OC@Nozieres3,OC@Demler.2013}%
. Using the trace formula~\cite{OC@Demler.2013}, we calculate the
radio-frequency spectrum of a half-filled one dimensional lattice at
zero-temperature, see Fig.\ref{fig3}. The spectrum exhibits in-gap peaks,
which imply bound states of the $\hat{c}$-fermions and local spins.

\begin{figure}[tbp]
\includegraphics[width=2.8in]{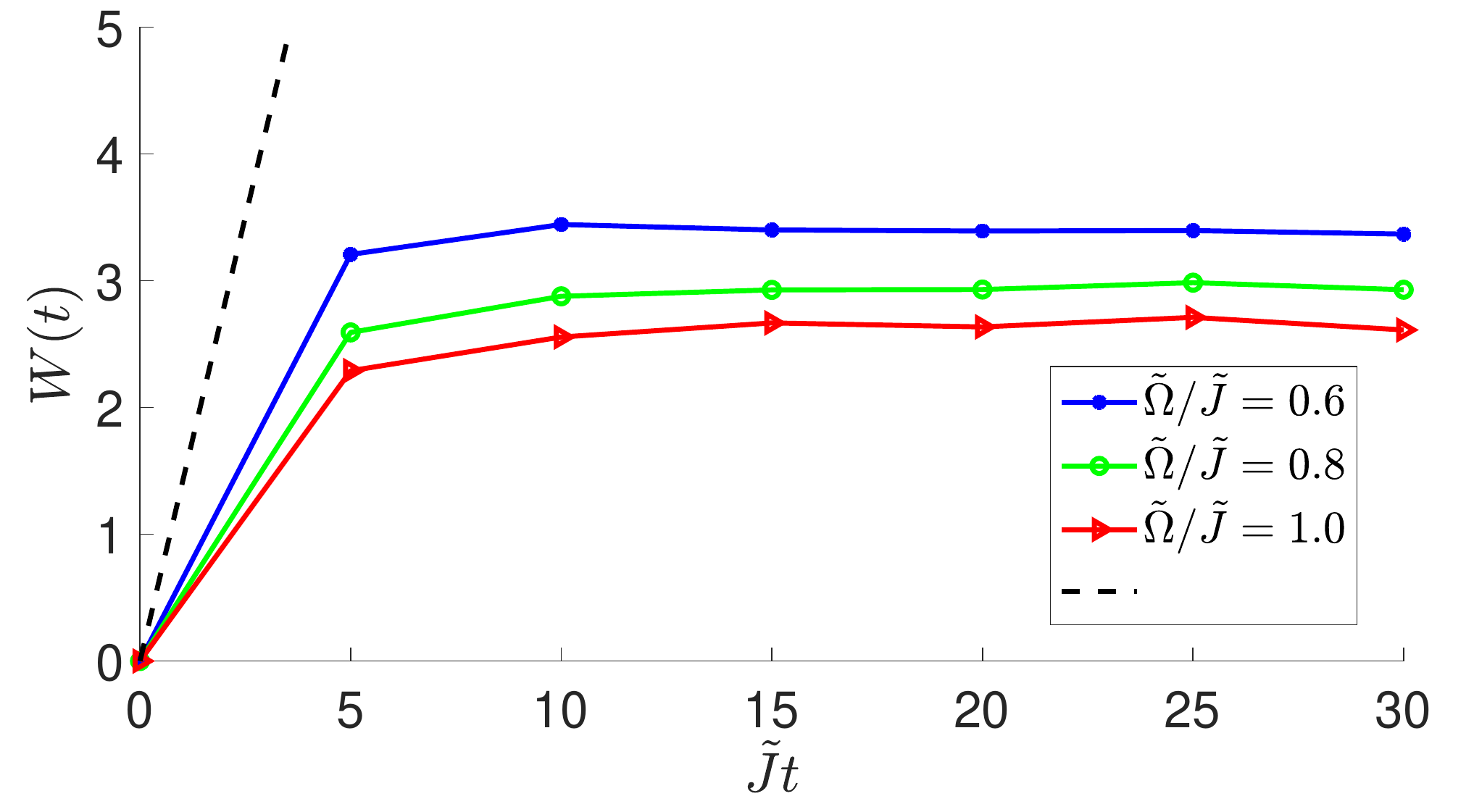}
\caption{Wavepacket spreading of one fermion in one dimension lattice gauge
field. The initial state of the solid lines is $\hat{U}^{\dag }\hat{c}_{%
\mathbf{0}}^{\dag }\left\vert 0\right\rangle \otimes \prod_{\mathbf{r}%
}\left\vert \uparrow \right\rangle _{\mathbf{r}}$, while the initial state
of the dashed line is $U^{\dag }\hat{c}_{\mathbf{0}}^{\dag }\left\vert
0\right\rangle \otimes \prod_{\mathbf{r}}\left\vert +1\right\rangle _{%
\mathbf{r}}$. The simulation is done on a lattice with length $21$. }
\label{fig4}
\end{figure}

Second observation is the dynamical localization. Consider the time
evolution of the initial states $\left\vert \psi (0)\right\rangle =\hat{U}%
^{\dag }\left\vert \varphi (0)\right\rangle \otimes \prod\limits_{\mathbf{r}%
}\left\vert \uparrow \right\rangle _{\mathbf{r}}$, where $\left\vert \varphi
(0)\right\rangle $ is the state of "Orthogonal" fermions, $\left\vert \uparrow
(\downarrow )\right\rangle _{\mathbf{r}}$ is the local eigenstate of $\sigma
_{\mathbf{r}}^{z}$. One can expand $\left\vert \uparrow \right\rangle _{%
\mathbf{r}}$ by the local eigenstates of $\hat{G}_{\mathbf{r}}^{\prime }$, $%
\left\vert \uparrow \right\rangle _{\mathbf{r}}=\frac{1}{2}\left( \left\vert
+1\right\rangle _{\mathbf{r}}+\left\vert -1\right\rangle _{\mathbf{r}%
}\right) $. So the initial state can be rewritten into $\left\vert \psi
(0)\right\rangle =\sqrt{2}^{-N_{\mathrm{s}}}\hat{U}^{\dag }\left\vert
\varphi (0)\right\rangle \otimes \sum_{\mathbf{g}}\left\vert \mathbf{g}%
\right\rangle $, where $N_{\mathrm{s}}$ is the number of lattice site. Note
that this wave function is a superposition of different subspaces. If we
focus on the behavior of the fermions, the expectation value of a fermion
operator reads%
\begin{equation}
\left\langle O\left( t\right) \right\rangle =\frac{1}{2^{N_{\mathrm{s}}}}%
\sum_{\mathbf{g}}\left\langle \varphi (0)\right\vert e^{\frac{iH_{\mathrm{sub%
}}\left( \mathbf{g}\right) t}{\hbar }}Oe^{-\frac{iH_{\mathrm{sub}}\left(
\mathbf{g}\right) t}{\hbar }}\left\vert \varphi (0)\right\rangle .
\end{equation}%
We can see that $e^{-\frac{iH_{\mathrm{sub}}\left( \mathbf{g}\right) t}{%
\hbar }}\left\vert \varphi (0)\right\rangle $ is the time evolution of the
non-interacting "Orthogonal" fermions in a subsector with given background
charge distribution. The summation is identical to the ensemble average of the "Orthogonal" fermions moving in disorder potential. So this dynamics is similar
to the Anderson model, expect the potential here is binary distribution
rather than the uniform distribution. In one or two dimension this dynamics
exhibits the localization~\cite%
{Localization@Moessner.2017,Localization@Scardicchio.2018,Localization@Kovrizhin.2018,Localization@Langari.2019,Localization@Zhai.2020,Localization@Knolle.2020}%
. However start with the initial state $\hat{U}^{\dag }\left\vert \varphi
(0)\right\rangle \otimes \prod_{\mathbf{r}}\left\vert +1\right\rangle _{%
\mathbf{r}}$, which is in one subsector. The "Orthogonal" fermions feel a
uniform potential, no background charge fluctuations. There will be no
localization. We calculate the spreading of a single fermion wavepacket in
one dimension, see Fig.\ref{fig4}. Starting from the uniform background
charges, the width of the wavepacket, $W=\sqrt{\left\langle \sum\nolimits_{%
\mathbf{r}}\mathbf{r}^{2}\hat{c}_{\mathbf{r}}^{\dag }\hat{c}_{\mathbf{r}%
}\right\rangle -\left\langle \sum\nolimits_{\mathbf{r}}\mathbf{r}\hat{c}_{%
\mathbf{r}}^{\dag }\hat{c}_{\mathbf{r}}\right\rangle ^{2}}$, grows linear
with the time. However if the initial state is a superposition state, the
width is saturated at long time. So the disorder effect in the dynamics is
led by the background charge fluctuation, which is the consequence of the
superposition initial states.

\textit{Summary.} We propose a Floquet approach to simulate a simple lattice
gauge model. Our method is feasible with current experiment techniques, and
can be implemented in extend system in one, two or three dimensions. The
effective Hamiltonian is obtained by keeping to the zeroth order of $%
1/\omega $ expansion. The next order processes with break the local gauge
symmetry. However, those terms are high suppressed in the large driving
frequency limit, such that the gauge invariant dynamics will not be affected
in the typical time scale of the experiment.

%\begin{figure}[]
%\includegraphics[width=3in]{fig1}

%\caption{ Caption of fig1 }

%\label{fig1}
%\end{figure}

\end{document}